\documentstyle[11pt,epsf]{article}
\def\bge{\begin{equation}}
\def\ene{\end{equation}}
\def\bg{\begin{eqnarray}}
\def\en{\end{eqnarray}}
\def\nn{\nonumber}

\begin{document}
\begin{center}
\vspace*{1.5cm}
{\bf \Large Hadron structure effect in finite nuclei }\\
\vspace*{5mm}
Koichi Saito \\
\vspace{3mm}
{\em
Physics Division, Tohoku College of Pharmacy, Sendai 981, Japan 
}
\end{center}
\vspace{.3cm}
%
%
The quark-meson coupling model, based on a mean-field description of 
non-overlapping nucleon bags bound by the self-consistent exchange of 
$\sigma$, $\omega$ and $\rho$ mesons, is reviewed.  
In particular, I present 
the changes in the hadron masses of 
the (non-strange) vector mesons and the nucleon in a nucleus.  
I also give a new, simple scaling relation among the changes of the hadron 
masses. 
%
%
\section*{\S 1. Introduction}
\vspace*{-3mm} \hspace*{3mm}

As the nuclear environment changes, hadron properties are nowadays expected to 
be modified~\cite{mass}.  
In particular, the variation of the light vector-meson mass 
is receiving a lot of attention, both theoretically and experimentally.  
Recent experiments from the HELIOS-3, CERES and NA50 
collaborations at SPS/CERN energies have shown that there exists a large 
excess of the lepton pairs in central heavy-ion 
collisions~\cite{qm96}.  On the other hand, 
even in low-energy nuclear physics, some attempts to measure the 
variation of hadron masses in nuclei are now underway~\cite{saka}.  

Recently we have developed the quark-meson coupling (QMC) model~\cite{sai1} 
to treat the variation of hadron properties in {\em finite\/} 
nuclei~\cite{sai2}. 
The QMC model may be viewed as an extension of QHD~\cite{qhd} in 
which the nucleons still interact through the exchange of scalar and 
vector mesons.  However, the mesons couple not to point-like nucleons but to
confined quarks (in the bag).  In studies of infinite nuclear 
matter~\cite{sai3} 
it was found that the quark degrees of freedom in the nucleon 
give an acceptable value for the 
incompressibility once the coupling constants are chosen to 
reproduce the correct saturation energy and density for symmetric 
nuclear matter. This is a significant improvement on QHD at the same level 
of sophistication.  

In this report I briefly review the QMC model and show some numerical 
results for static properties and variation of the hadron masses in 
some nuclei~\cite{sai2}. 
I also give a new, simple scaling relation among the changes of the 
hadron masses in a nucleus. 

\section*{\S 2. The quark-meson coupling model}
\vspace*{-3mm} \hspace*{3mm}
Let us suppose that a free nucleon (at the origin) consists of three light 
(u and d) quarks under a (Lorentz scalar) confinement potential, $V_c$.  
Then, the Dirac equation for the quark field, $\psi_q$, is given by 
\bge
[ i\gamma\cdot\partial - m_q - V_c(r) ] \psi_q(r) = 0 , 
\label{dirac1}
\ene
where $m_q$ is the bare quark mass.  

Next we consider how Eq.(\ref{dirac1}) is modified when the nucleon is bound 
in static, uniformly distributed (iso-symmetric) nuclear matter.  
In the QMC model it is assumed that 
each quark feels scalar, $V_s^q$, and vector, $V_v^q$, potentials, which are 
generated by the
surrounding nucleons, as well as the confinement potential. 
Since the typical distance between two nucleons around normal nuclear density 
is surely larger than the typical size of the 
nucleon (its radius $R_N \sim$ 0.8 fm), the interaction (except for 
the short-range part) between the nucleons 
should be colour singlet, e.g., a meson-exchange potential.  Therefore, this 
assumption seems appropriate when the baryon density, $\rho_B$, is not high.  
If we use the mean-field approximation for the meson fields, 
Eq.(\ref{dirac1}) may be rewritten as 
\bge
[ i\gamma\cdot\partial - (m_q - V_s^q) - V_c 
 - \gamma_0 V_v^q ] \psi_q = 0 . 
\label{dirac2}
\ene
The potentials generated by the medium are constants because the matter 
distributes uniformly. As the nucleon is static, the time-derivative 
operator can be replaced by the quark energy, $-i \epsilon_q$.  
By analogy with the procedure applied to the nucleon
in QHD, if we introduce the 
effective quark mass by $m_q^{\star} = m_q - V_s^q$, the Dirac equation, 
Eq.(\ref{dirac2}), can be rewritten in the same form as that in free space, 
with the mass $m_q^{\star}$ and the energy $\epsilon_q - V_v^q$, instead of 
$m_q$ and $\epsilon_q$. 
In other words, the vector interaction has {\em no effect 
on the nucleon structure} except for an overall phase in the quark wave 
function, which gives a shift in the nucleon energy.  This fact 
{\em does not\/} depend on how to choose the confinement potential, $V_c$.  
Then, the nucleon energy (at rest), $E_N$, in the medium is given by 
$E_N = M_N^{\star}(V_s^q) + 3V_v^q $, where the effective nucleon mass, 
$M_N^{\star}$, depends on {\em only the 
scalar potential}~\cite{sai3}. 

Now we extend this idea to finite nuclei. 
The solution of the general problem of a composite, quantum particle 
moving in background scalar and vector fields that vary with position is 
extremely difficult.  One has, however, a chance to solve the particular 
problem of interest to us, namely light quarks confined in a nucleon which is 
itself bound in a finite nucleus, only because the nucleon motion is 
relatively slow and the quarks highly relativistic~\cite{sai1}. Thus the 
Born-Oppenheimer approximation is naturally suited to 
the problem. 

Our approach in Ref.\cite{sai1} was to start with a classical nucleon 
and to allow its internal structure 
to adjust to minimise the energy of three quarks in the ground-state of 
a system under constant scalar and vector fields, 
with values equal to those at the centre of the nucleon. 
In our case, the MIT bag model was used to describe the nucleon 
structure (see also Ref.~\cite{blun}). 
Of course, the major problem with 
the MIT bag (as with many other relativistic models of nucleon structure) 
is that it is difficult to boost. We therefore solve the bag 
equations in the instantaneous rest frame (IRF) of the nucleon -- using a
standard Lorentz transformation to find the energy and momentum of the
classical nucleon bag in the nuclear rest frame. 
Having solved the problem using the meson fields at the centre of 
the nucleon, 
one can use perturbation theory to correct for the variation of the 
scalar and vector fields across the nucleon bag. In first order perturbation 
theory only the spatial components of the vector potential 
give a non-vanishing contribution. 
This extra term is a correction to the spin-orbit force~\cite{sai1}. 

As shown in Refs.\cite{sai1,sai2}, the basic result in the QMC model 
is that, in the scalar and vector fields, 
the nucleon behaves essentially as a point-like 
particle with an effective mass 
$M_N^{\star}$, which depends on the position through only the scalar 
field, moving in the vector potential. 

Let us consider that the scalar and vector potentials in Eq.(\ref{dirac2}) are 
mediated by the $\sigma$ and $\omega$ mesons, and introduce their 
mean-field values, which now depend on position ${\vec r}$, 
by $V_s^q({\vec r}) = g_{\sigma}^q \sigma({\vec r})$ and 
$V_v^q({\vec r}) = g_{\omega}^q \omega({\vec r})$, respectively, where 
$g_{\sigma}^q$ ($g_{\omega}^q$) is the coupling constant of the quark-$\sigma$ 
($\omega$) meson.  Furthermore, we shall add the isovector, vector meson, 
$\rho$, and the Coulomb field, $A({\vec r})$, to describe finite nuclei 
realistically~\cite{sai1,sai2}.  
Then, the effective lagrangian density for finite nuclei, involving the quark 
degrees of freedom in the nucleon and the (structureless) meson fields 
in MFA, would be given by 
\bg
{\cal L}_{QMC-I}&=& \overline{\psi} \left[ i \gamma \cdot \partial 
- M_N^{\star} - g_\omega \omega \gamma_0 
- \frac{g_\rho}{2} \tau^N_3 b \gamma_0 
- \frac{e}{2} (1+\tau^N_3) A \gamma_0 \right] \psi \nn \\
&-& \frac{1}{2}[ (\nabla \sigma)^2 + m_{\sigma}^2 \sigma^2 ] 
+ \frac{1}{2}[ (\nabla \omega)^2 + m_{\omega}^2 \omega^2 ] 
+ \frac{1}{2}[ (\nabla b)^2 + m_{\rho}^2 b^2 ] 
+ \frac{1}{2} (\nabla A)^2 , 
\label{qmclag}
\label{qmc-1}
\en
where $\psi$ and $b$ are respectively the 
nucleon and the $\rho$ fields. 
$m_\sigma$, $m_\omega$ and $m_{\rho}$ are respectively 
the (constant) masses of the $\sigma$, $\omega$ and $\rho$ mesons. 
$g_\omega$ and $g_{\rho}$ are respectively the $\omega$-N and $\rho$-N 
coupling constants, which are given by 
$g_\omega = 3 g_\omega^q$ and $g_\rho = g_\rho^q$ (where 
$g_\rho^q$ is the quark-$\rho$ coupling constant).  
We call this model the QMC-I model.  
If we define the field-dependent $\sigma$-N coupling 
constant, $g_\sigma(\sigma)$, by
\bge
M_N^{\star}(\sigma({\vec r})) \equiv M_N - g_\sigma(\sigma({\vec r})) 
\sigma({\vec r}) , \label{coup}
\ene
where $M_N$ is the free nucleon mass, it is easy to compare with 
QHD.  
The difference between QMC-I  
and QHD lies only in the coupling constant $g_\sigma$, which
depends on the scalar field in QMC-I while it is constant in QHD.  
However, this difference leads to a lot of favorable results, notably the 
nuclear incompressibility~\cite{sai1}. 

Here we consider the nucleon mass in matter further.  The nucleon mass is a 
function of the scalar field.  Because the scalar field is small 
at low density it can be expanded in terms of $\sigma$ as 
\bge
M_N^{\star} = M_N + \left( \frac{\partial M_N^{\star}}{\partial \sigma} 
\right)_{\sigma=0} \sigma + \frac{1}{2} \left( \frac{\partial^2 M_N^{\star}}
{\partial \sigma^2} \right)_{\sigma=0} \sigma^2 + \cdots . 
\label{nuclm}
\ene
Since the interaction Hamiltonian between the nucleon and the 
$\sigma$ field at the quark level is given by $H_{int} = - 3 g_{\sigma}^q 
\int d{\vec r} \ \overline{\psi}_q \sigma \psi_q$, the derivative of 
$M_N^{\star}$ with respect to $\sigma$ is given by 
$ -3g_{\sigma}^q \int d{\vec r} \ \ {\overline \psi}_q \psi_q 
(\equiv -3g_{\sigma}^q S_N(\sigma))$, where 
we define the quark-scalar density in the nucleon, $S_N(\sigma)$,
which is itself a function of the scalar field. 
Because of a negative value of the derivative 
the nucleon mass decreases in matter at low density.  

Furthermore, we define the scalar-density ratio, $S_N(\sigma)/S_N(0)$, 
to be $C_N(\sigma)$ and the $\sigma$-N coupling constant at $\sigma = 0$  
to be $g_\sigma$ (i.e., $C_N(\sigma) = S_N(\sigma)/S_N(0)$ and 
$g_{\sigma} = g_\sigma(\sigma=0) = 3g_{\sigma}^q S_N(0)$). 
Using these quantities, the nucleon mass is rewritten as 
\bge
M_N^{\star} = M_N - g_{\sigma} \sigma - \frac{1}{2} g_{\sigma} 
C_N^\prime(0) \sigma^2 + \cdots . 
\label{nuclm2}
\ene
In general, $C_N$ is a decreasing function because the quark in matter is 
more relativistic than in free space.  Thus, $C_N^\prime(0)$ takes a 
negative value. If the nucleon were structureless $C_N$ would not depend on 
the scalar field, that is, $C_N$ would be constant ($C_N=1$).  Therefore, 
only the first two terms in the right hand side of Eq.(\ref{nuclm2}) remain, 
which is exactly the same as the equation for the effective nucleon 
mass in QHD (see also Ref.~\cite{wallace,sai35}).  
By taking the heavy-quark-mass limit in QMC we can reproduce 
the QHD results~\cite{sai1,sai2,sai3}.  

%
%
%
%
%
%
%

We have considered the effect of nucleon 
structure.  It is however true that the mesons are also built of quarks and 
anti-quarks, and that they may change their properties in matter.  
To incorporate the effect of meson structure in the QMC model, we 
suppose that the vector mesons are again described by a relativistic quark 
model with {\em common\/} scalar and vector mean-fields~\cite{sai2}, 
like the nucleon (see Eq.(\ref{dirac2})).  
Then, again the effective vector-meson mass in matter, 
$m_v^{\star}$ ($v$ = $\omega$ or $\rho$), depends on only the 
scalar mean-field.  

However, for the scalar ($\sigma$) meson it may not be easy to describe it 
by a simple quark model (like a bag) because it couples strongly 
to the pseudoscalar ($2 \pi$) channel, which requires a direct 
treatment of chiral symmetry in medium~\cite{hatkun}.  Since, according to the 
Nambu--Jona-Lasinio model or the Walecka model, 
one might expect the 
$\sigma$-meson mass in medium, $m_{\sigma}^{\star}$, to be less than the free 
one, we shall here parametrize it using a quadratic 
function of the scalar field (by hand)~\cite{sai2}: 
\bge
\left( \frac{m_{\sigma}^{\star}}{m_{\sigma}} \right) = 1 - a_{\sigma} 
(g_{\sigma} \sigma) + b_{\sigma} (g_{\sigma} \sigma)^2 , 
\label{sigmas}
\ene
with $g_{\sigma} \sigma$ in MeV, and we introduce two parameters, 
$a_{\sigma}$ (in MeV$^{-1}$) and $b_{\sigma}$ (in MeV$^{-2}$).  
We will determine these parameters later.  

Using these effective meson masses, we can find a new lagrangian 
density for finite nuclei, 
which involves the structure effects of both the nucleon and 
mesons in MFA: 
\bg
{\cal L}_{QMC-II}&=& \overline{\psi} \left[ i \gamma \cdot \partial 
- M_N^{\star} - g_\omega \omega \gamma_0 
- \frac{g_\rho}{2} \tau^N_3 b \gamma_0 
- \frac{e}{2} (1+\tau^N_3) A \gamma_0 \right] \psi \label{qmc-2} \\
&-& \frac{1}{2}[ (\nabla \sigma)^2 + 
m_{\sigma}^{\star 2} \sigma^2 ] 
+ \frac{1}{2}[ (\nabla \omega)^2 + m_{\omega}^{\star 2} 
\omega^2 ] + \frac{1}{2}[ (\nabla b)^2 + m_{\rho}^{\star 2} b^2 ] 
+ \frac{1}{2} (\nabla A)^2 . \nn 
\en
We call this model QMC-II.  
Note that all the masses depend on the scalar mean-field.  

%
%

%
\section*{\S 3. Numerical results}
\vspace*{-3mm} \hspace*{3mm}

In this section, we will show our numerical results using the QMC-II model. 
(For QMC-I, see Ref.~\cite{sai1}.)

For infinite nuclear matter we take the Fermi momenta for protons and 
neutrons to be $k_{F_i}$ ($i=p$ or $n$). This is 
defined by $\rho_i = k_{F_i}^3 / (3\pi^2)$, where $\rho_i$ is the 
density of protons or neutrons, and the total baryon density, 
$\rho_B$, is then given by $\rho_p + \rho_n$.  
(Note that the mean-field values for the mesons are {\em constant}.) 

From the lagrangian density Eq.(\ref{qmc-2}), we can easily find the 
total energy per nucleon, $E_{tot}/A$, and the values of the $\omega$ and 
$\rho$ fields 
(which are respectively given by baryon number conservation and 
the difference in proton and neutron densities, $\rho_3 = \rho_p - \rho_n$). 
On the other hand, the scalar mean-field is given by a self-consistency 
condition (SCC): $\left( \frac{dE_{tot}}{d \sigma} \right) =0$.  

Now we need a model for the structure of the hadrons involved.  We use the MIT 
bag model in static, spherical cavity approximation~\cite{MIT}.  
In the model, the bag constant $B$ and the 
parameter $z_N$ in the familiar form of the 
MIT bag model lagrangian are fixed to reproduce the free nucleon mass 
($M_N$ = 939 MeV) under the  
condition that the hadron mass be stationary under variation of
the free bag radius ($R_N$ in the case of the nucleon).  
Furthermore, to fit the free vector-meson masses, 
$m_{\omega}$ = 783 MeV and $m_{\rho}$ = 770 MeV, we introduce new 
$z$-parameters for them, $z_{\omega}$ and $z_{\rho}$.  
In the following we choose $R_N=0.8$ fm and the free quark mass 
$m_q$ = 5 MeV.  
We then find that $B^{1/4}$ = 170.0 MeV, $z_N$ = 3.295, 
$z_\omega$ = 1.907 and $z_\rho$ = 1.857.  
(Variations of the quark mass and $R_N$ only lead to 
numerically small changes in the calculated results~\cite{sai1,sai2}.)  

Next we must choose the two parameters in the parametrization for the 
$\sigma$-meson mass in matter (see Eq.(\ref{sigmas})).  Here we 
consider three parameter sets: 
(A) $a_\sigma = 3.0 \times 10^{-4}$ (MeV$^{-1}$) and 
$b_\sigma = 100 \times 10^{-8}$ (MeV$^{-2}$), 
(B) $a_\sigma = 5.0 \times 10^{-4}$ (MeV$^{-1}$) and 
$b_\sigma = 50 \times 10^{-8}$ (MeV$^{-2}$), 
(C) $a_\sigma = 7.5 \times 10^{-4}$ (MeV$^{-1}$) and 
$b_\sigma = 100 \times 10^{-8}$ (MeV$^{-2}$).  The parameter sets A, B and 
C give about 2\%, 7\% and 10\% decreases of the $\sigma$ mass 
at saturation density, respectively~\cite{sai2}.

Now we are in a position to determine the coupling constants.  
$g_{\sigma}$ and $g_{\omega}$ are fixed to fit the binding energy 
($-15.7$ MeV) at the saturation density ($\rho_0 = 0.15$ fm$^{-3}$) for 
symmetric nuclear matter.  Furthermore, the $g_\rho$ is 
used to reproduce the bulk symmetry energy, 35 MeV.  
We take $m_\sigma$ = 550 MeV. The coupling constants 
and some calculated properties for matter in QMC-II 
are listed in Table~\ref{ccc}.  
The last three columns show the relative changes (from their values at 
zero density) of the nucleon-bag radius 
($\delta R_N^{\star}/R_N$), the lowest eigenvalue ($\delta x_N^{\star}/x_N$) 
and the root-mean-square radius (rms radius) of the nucleon calculated using 
the quark wave function ($\delta r_q^{\star}/r_q$) at saturation density. 
\begin{table}[htbp]
\begin{center}
\caption{Coupling constants and calculated properties for 
symmetric nuclear matter at normal nuclear density ($m_q$ = 5 MeV, 
$R_N$ = 0.8 fm and $m_\sigma$ = 550 MeV).  
The effective nucleon mass, $M_N^{\star}$, and the nuclear 
incompressibility, $K$, are quoted in MeV. 
The bottom row is for QHD.}
\label{ccc}
\begin{tabular}[t]{c|cccccccc}
\hline
type & $g_{\sigma}^2/4\pi$&$g_{\omega}^2/4\pi$&$g_\rho^2/4\pi$&
$M_N^{\star}$&$K$&$\delta R_N^{\star}/R_N$&$\delta x_N^{\star}/x_N$&$\delta 
r_q^{\star}/r_q$ \\
\hline
 A &  3.84 & 2.70 & 5.54 & 801 & 325 & $-0.01$ & $-0.11$ & 0.02 \\
 B &  3.94 & 3.17 & 5.27 & 781 & 382 & $-0.01$ & $-0.13$ & 0.02 \\
 C &  3.84 & 3.31 & 5.18 & 775 & 433 & $-0.02$ & $-0.14$ & 0.02 \\
\hline
 QHD & 7.29 & 10.8 & 2.93 & 522 & 540 & --- & --- & --- \\
\hline
\end{tabular}
\end{center}
\end{table}
We note that the nuclear incompressibility is higher than that in QMC-I 
($K \sim$ 200 -- 300 MeV)~\cite{sai1,sai3}.  However, it is still much 
lower than in QHD.  As in QMC-I, the bag radius of the 
nucleon shrinks a little, while its rms radius swells a little.  On the 
other hand, because of the $\sigma$ field, the eigenvalue is reduced more 
than 10\% (at $\rho_0$) from that in free space. 

In the present model it is possible to calculate masses of other hadrons  
like hyperons in medium~\cite{sai2}  
(see also Ref.~\cite{sai4}.) In that case, 
we assume that for the hyperons themselves we again use the MIT bag model,   
and that the strange quark in the hyperon does not directly couple 
to the scalar field in MFA, as one would expect if the $\sigma$ meson
represented a two-pion-exchange potential. It is also assumed
that the addition of a single hyperon to nuclear matter of density 
$\rho_B$ does not alter the values of the scalar and vector mean-fields, 
namely, we take the local-density approximation to the hyperons.  

In general, as in the nucleon case (see Eq.(\ref{nuclm})), we can expand 
the effective hadron mass in nuclear matter at low $\rho_B$ as 
\bg
M_j^{\star} &=& M_j + \left( \frac{\partial M_j^{\star}}{\partial \sigma} 
\right)_{\sigma=0} \sigma + \frac{1}{2} \left( \frac{\partial^2 M_j^{\star}}
{\partial \sigma^2} \right)_{\sigma=0} \sigma^2 + \cdots , \nn \\
  &\simeq& M_j - \frac{n_0}{3} g_\sigma \sigma 
  - \frac{n_0}{6} g_\sigma C_j^\prime(0) \sigma^2 , 
\label{hm}
\en
where $j$ stands for $N$, $\omega$, $\rho$, $\Lambda$, $\Sigma$, $\Xi$, etc., 
$n_0$ is the number of non-strange quarks in the hadron $j$ 
and $C_j(\sigma) (= S_j(\sigma)/S_j(0))$ is the scalar-density ratio 
defined by the quark-scalar density, $S_j$, in $j$.  
Since the strength of the scalar mean-field, $g_{\sigma} \sigma$, 
at low $\rho_B$ can be approximated by a linear function of the 
density~\cite{sai2} as $g_{\sigma} \sigma \approx $ 200 (MeV) 
$(\rho_B/\rho_0)$, 
we find the hadron mass at low $\rho_B$: 
\bge
\left( \frac{M_j^{\star}}{M_j} \right) \simeq 1 - 
c_j \left( \frac{\rho_B}{\rho_0} \right), 
\label{nuclvect}
\ene
where $c_{N, v, \Lambda, \Sigma, \Xi} = 0.21, 0.17, 0.12, 0.11, 0.05$, 
respectively .  

Furthermore, using the MIT bag model 
we find that the scalar-density ratio $C_j$ can be well approximated by 
a linear function of the scalar field~\cite{sai2}: 
$C_j(\sigma) = 1 - a_j \times (g_{\sigma} \sigma)$, 
where $a_j$ is the slope parameter for the hadron $j$, and that 
the dependence of $a_j$ on the 
hadrons is quite weak (it ranges around $8.6 \sim 9.5 \times 
10^{-4}$ (MeV$^{-1}$)).  
Therefore, 
if we ignore its weak dependence on the hadrons, the effective hadron 
mass can be rewritten in a quite simple form (from Eq.(\ref{hm})): 
\bge
M_j^{\star} \simeq M_j - \frac{n_0}{3} (g_\sigma \sigma) \left[ 1 - 
  \frac{a}{2} (g_\sigma \sigma) \right] , 
\label{hm3}
\ene
where $a \simeq 9.0 \times 10^{-4}$ (MeV$^{-1}$).  This mass formula can 
reproduce the hadron masses in 
matter quite well over a wide range of $\rho_B$ (up to $\sim 3 \rho_0$). 
Therefore, once one knows the behaviour of $g_\sigma \sigma$ at finite 
density, one can easily calculate the effective hadron mass 
using Eq.(\ref{hm3}). 
A simple parametrization of $g_{\sigma} \sigma$ is given in Ref.~\cite{sai5}. 

Since the scalar field is common to all hadrons, Eq.(\ref{hm3}) leads to 
a new, simple scaling relationship among the hadron masses~\cite{sai2}: 
\bge
\left( \frac{\delta m_v^{\star}}{\delta M_N^{\star}} \right) \simeq 
\left( \frac{\delta M_\Lambda^{\star}}{\delta M_N^{\star}} \right) \simeq 
\left( \frac{\delta M_\Sigma^{\star}}{\delta M_N^{\star}} \right) \simeq 
\frac{2}{3} \ \ \ \mbox{ and } \ \ \ 
\left( \frac{\delta M_\Xi^{\star}}{\delta M_N^{\star}} \right) \simeq 
\frac{1}{3} , 
\label{scale}
\ene
where $\delta M_j^{\star} \equiv M_j - M_j^{\star}$.  The factors, 
$\frac{2}{3}$ 
and $\frac{1}{3}$, in Eq.(\ref{scale}) come from the ratio of the number of 
non-strange quarks in $j$ to that in the nucleon.  
It would be very interesting to see whether this scaling relationship is 
correct in forthcoming experiments.  

\begin{figure}[h,t]
\begin{center}\leavevmode
\epsfysize=7cm
\epsfbox{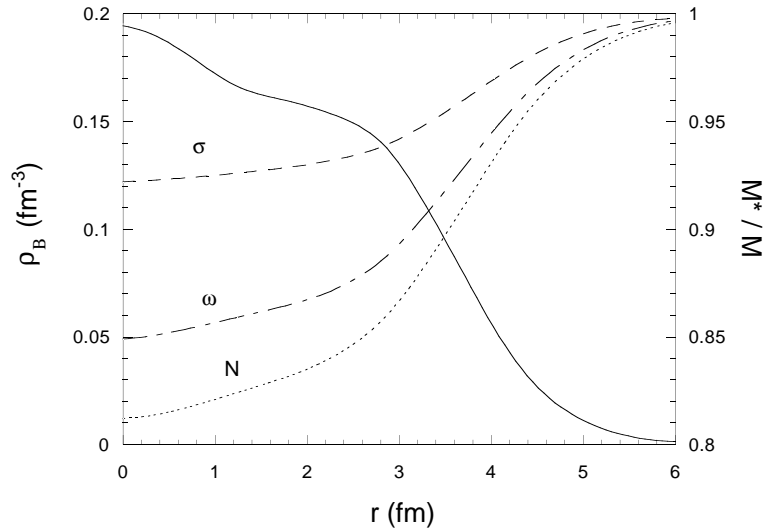}
\caption{Changes of the nucleon, $\sigma$ and $\omega$ meson masses 
in $^{40}$Ca.  The nuclear baryon density is also illustrated (solid curve).  
The right (left) scale is for the effective mass (the baryon density). 
The parameter set B is used.  }
\label{hmca40}
\end{center}
\end{figure}
For finite nuclei, we will show our results of some finite, closed shell 
nuclei. We have solved self-consistently a set of 
coupled non-linear differential equations~\cite{num} given by 
the lagrangian density Eq.(\ref{qmc-2}).  
In Fig.~\ref{hmca40} we present our self-consistent, full 
results of the changes of the 
nucleon, $\sigma$ and $\omega$ meson masses in $^{40}$Ca.  
%
%
Table~\ref{sum} gives a summary of the calculated binding energy per nucleon 
($E/A$), rms charge radii and the difference between nuclear rms radii for 
neutrons and protons ($r_n - r_p$) for several closed-shell 
nuclei~\cite{sai1,sai2}.  
\begin{table}[htbp]
\begin{center}
\caption{Binding energy per nucleon, $- E/A$ (in MeV), rms charge radius 
$r_{ch}$ (in fm) and the difference between $r_n$ and $r_p$ (in fm) for 
type B, 
$m_q$ = 5 MeV and $R_B$ = 0.8 fm. 
I and II denote, respectively, QMC-I and QMC-II. ($^*$ fit) }
\label{sum}
\begin{tabular}[t]{c|ccc|ccc|ccc}
\hline
 & & $-E/A$ & & & $r_{ch}$ & & & $r_n-r_p$ & \\
\hline
model&I&II&Expt.&I&II&Expt.&I&II& Expt. \\
\hline
$^{16}$O &5.84&5.11&7.98&2.79&2.77&2.73&$-0.03$&$-0.03$&0.0 \\
$^{40}$Ca&7.36&6.54&8.45&3.48$^*$&3.48$^*$&3.48&$-0.05$&$-0.05$&0.05$\pm$0.05\\
$^{48}$Ca&7.26&6.27&8.57&3.52&3.53&3.47&0.23&0.24&0.2$\pm$0.05 \\
$^{90}$Zr&7.79&6.99&8.66&4.27&4.28&4.27&0.11&0.12&0.05$\pm$0.1 \\
$^{208}$Pb&7.25&6.52&7.86&5.49&5.49&5.50&0.26&0.27&0.16$\pm$0.05 \\
\hline
\end{tabular}
\end{center}
\end{table}
While there are still some discrepancies between 
the results and data, the present model provides reasonable results.  
In particular, as in QMC-I, it reproduces the rms charge radii for medium 
and heavy nuclei quite well.  

%
\section*{\S 4. Summary}
\vspace*{-3mm} \hspace*{3mm}

I have reviewed the quark-meson coupling (QMC) model to include  
quark degrees of freedom in the hadrons involved, and have shown 
the density dependence of hadron masses in $^{40}$Ca as an example.  
As several authors have 
suggested~\cite{mass,qm96}, the hadron mass is 
reduced because of the scalar mean-field in medium.  
In the present model the hadron mass can be related to the number of 
non-strange quarks and the strength of the scalar mean-field 
(see Eq.(\ref{hm3})).  
We have found a new, simple formula to describe the hadron masses in the 
medium, and this led to a new scaling relationship among them 
(see Eq.(\ref{scale})).  
We should note that the origins of the meson-mass reduction in QMC and 
QHD are completely different~\cite{sai2,sai35}.  
It would be very interesting to compare our results with forthcoming 
experiments on the vector-meson mass~\cite{saka}.  

At low energy an effective field theory to describe nuclei will 
generally contain an infinite 
number of interaction terms, which incorporate the {\em compositeness\/} of 
the low-energy degrees of freedom, namely the hadrons~\cite{serot2}, and it 
is then expected to involve numerous couplings which may be nonrenormalizable. 
Thus, one needs an {\em organizing principle\/} to find out a sensible 
field theory.  Manohar and Georgi~\cite{nda} have proposed a systematic way to 
manage such complicated, effective field theories called ``naive
dimensional analysis'' (NDA).
NDA gives rules for assigning a coefficient of the 
appropriate size to any interaction term in an effective lagrangian.  
After extracting the dimensional factors and 
some appropriate counting factors using NDA, the remaining 
{\em dimensionless\/} 
coefficients are all assumed to be of order {\em unity}.  This is the 
so-called {\em naturalness assumption}.  

Our lagrangian density, Eq.(\ref{qmc-2}), provides a lot of effective 
coupling terms among the meson fields because the mesons have structure.  
In particular, the lagrangian 
automatically offers self-coupling terms (or non-linear terms) with 
respect to the $\sigma$ field.  Therefore, it is very interesting whether 
the QMC model gives {\em natural\/} coupling constants.  In Ref.~\cite{sai6}, 
we examined the QMC model using NDA, and concluded that the QMC model 
is quite {\em natural\/} as an effective field theory for nuclei.  

\vspace{0.5cm}
The author thanks Tony Thomas and Kazuo Tsushima for valuable discussions and 
comments. 
%

%
\end{document}